\documentclass[useAMS]{mn2e}
\usepackage{graphicx}

\title [Optical Photopolarimetry of Blazar OJ287]{Optical Photopolarimetry of Blazar OJ287 \thanks{Based on data collected with 2-m RCC telescope with FoReRo2 at Rozhen National Astronomical Observatory, Bulgaria.}}

\author[V. Bozhilov et al.]{V. Bozhilov$^{1}$\thanks{vbozhilov@phys.uni-sofia.bg}, E. Ovcharov$^{1}$ and G. Nikolov$^{2}$ \\
$^{1}$Department of Astronomy, Sofia University "St. Kliment Ohridski", 5 James Bourchier Blvd., BG--1164 Sofia, Bulgaria.\\
$^{2}$Institute of Astronomy and National Astronomical Observatory, Bulgarian Academy of Science, 72, Tsarigradsko Chaussee Blvd. 1784, Sofia}

\begin{document}

\date{Accepted 2013 December 23.  Received 2013 December 23; in original form 2013 November 12}

\pagerange{\pageref{firstpage}--\pageref{lastpage}} \pubyear{2013}

\maketitle

\label{firstpage}

\begin{abstract}
We present results from an original observational campaign comprising five epoch optical photopolarimetrical observations of the BL Lac-type AGN OJ287 in the period 2012 November - 2013 April. The data are gathered with the Focal Reducer Rozhen 2 -- FoReRo2 on the 2-m RCC telescope at NAO Rozhen, Bulgaria. We derive photometry and polarization in R-band, as well as position angle (P.A.). There are indications for correlation between polarization and brightness in R-band. Furthermore, observed variation in P.A. corresponds to a rotation of the plane of polarization of 5.80 $deg$ per day.
\end{abstract}


\section{Introduction}
Blazars are very powerful and extremely variable sources of polarized radiation. At z=0.306, the BL Lac-type object OJ287  is one of the most well-observed blazars. In fact, observations on its light curve date back to year 1891 (Valtonen \& Ciprini 2012). But OJ287 is one of the most peculiar AGNs, as well. It shows well-defined period of 12 years (Sillanp\"{a}\"{a} et al. 1996). Different theoretical models has been proposed to explain its characteristically double-peaked flaring activity during outburst (Katz 1997; Villata et al. 1998; Sillanp\"{a}\"{a} et al. 1988; Sundelius et al. 1997; Valtaoja et al. 2000). The favored current  model is of a binary black hole system with relativistic precession (Valtonen \& Ciprini 2012). It is interesting to note that, according to Neronov \& Vovk (2011), the relativistic beamed emission may come from the jet, produced by the smaller secondary black hole.

The analysis of polarization behaviour plays a key role in the study of blazars (Barres de Almeida et al. 2010). Polarization measurements for OJ287, especially during outburst (Valtaoja et al. 2000), are of crucial importance. The motivation is that the first of the two flares in OJ287 during outburst is thermal (i.e. not polarized). This is shown by the lack of corresponding radio emission (Valtaoja et al. 2000). But the second flare is model-dependent and could be due either to synchrotron radiation, which is polarized (Valtaoja et al. 2000), or can be unpolarized or low-polarized bremsstrahlung (Valtonen \& Ciprini 2012). The polarization is a key factor to distinguish among different theoretical models.

A number of multiwavelength observational campaigns have been performed in the last years (Valtaoja et al. 2000; Efimov et al. 2002; Ciprini et al. 2007; Nieppola et al. 2009; Villforth et al. 2010; Valtonen \& Ciprini 2012). Results from the 2005-2010 observational campaign show that the flare during the 2005 outburst is due to bremsstrahlung radiation instead of a synchrotron one (Valtonen \& Ciprini 2012). That provides strong background to the binary black hole model with ultra relativistic precession. This also allows for the use of OJ287 as a test of general relativity (Valtonen \& Ciprini 2012). Surprisingly, in April 2012 OJ287 showed a peak in brightness (Santangelo 2012) that was not expected by any of the current models. Possible theoretical explanations according to the binary black hole model can be found in Pihajoki et al. (2013).

There are indications that position angle (P.A.) change is between 2 $deg$ (D'arcangelo et al. 2009) and 5 $deg$ per day (Efimov et al. 2002). But there is a real shortage of optical photopolarimetrical campaigns. Possible correlation between optical flux and polarization degree for OJ287 is suggested by observations from Takalo et al. 1994, but is not well estabished so far (Jorstad et al. 2007, Villforth et al. 2010).

Complete data of polarization behaviour and P.A. measurements, combined with simultaneous photometrical data of OJ287, are important for understanding the underlying physics of this peculiar object. But our work is not aimed at discussing the various theoretical models for OJ287. Instead, we focus on original observational photopolarimetrical study and, as a result, we get some exciting new insights.


\section{Observations and Data Reduction}

We present original photopolarimetrical study of OJ287 with the Focal Reducer Rozhen 2 -- FoReRo2 (Jockers et al. 2000) on the 2-m Ritchey-Cretien-Coude (RCC) telescope at NAO Rozhen, Bulgaria. The observational data are taken on the nights of 2012 November 17 and 18 (preliminary results of these two nights are published in Bozhilov, Borisov \& Ovcharov (2013)), 2013 January 13, 2013 Arpil 04 and 05. Data from the last three nights are published for the first time, as well is the full-scale analysis of our observational campaign on OJ287.  

\begin{figure} 
      \centering
      \includegraphics[width=40mm,height=40mm]{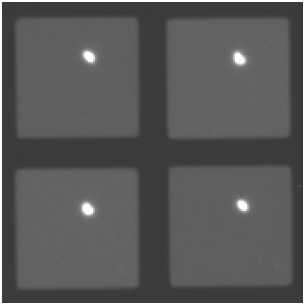}
      \caption{
CCD-image from the polarizer with two combined Wollaston prisms. Each of the four squares correspond to a polarized beam orientations at (respectively) 0$deg$, 90$deg$, 45$deg$ and 135$deg$.
              }
      \label{fig4}
   \end{figure}

We use color splitter that transmits redder than 5800\AA\ light into the red channel and reflects bluer than 5100\AA\ one into the blue channel of the reducer. Polarimetrical measurements in B are not performed, but B-band data are used for photometrical measurements (Bozhilov et al. 2013). We used a special optical element with two Wollaston prisms. They are combined to form a single polarizer. The P.A. of the prisms differ by 45 $deg$. Thus, we get four polarized beams with orientations at 0 $deg$, 90 $deg$, 45 $deg$ and 135 $deg$ each. These are simultaneously captured by the CCD detector. On Figure \ref{fig4} is shown an example of real-life image.

Armed with this experimental setup, we can perform polarization determination using Stokes equations (see  Landi Degl'Innocenti \& M. Landolfi 2004, chapter 1.6 and 1.7 for elaborate details). A detailed description of the method of measurements and the calculations can be found in Geyer et al. (1996). 

Details on the observations of OJ287 and standard stars, number of images, exposures and total integration time are presented on Table \ref{tab1}.

In order to determine instrumental optical polarization, according to Stokes equations (Geyer et al. 1996) we used low-polarization standard stars HD10476, HD90508, HD144287  and high-polarization standard stars HD14433, HD43384, HD154445 for the nights of November 2012, January 2013 and April 2013 respectively (see Table \ref{tab2}). Here we have to note that the deviation of the measured P.A. for the high-polarized stars are relatively high. Thus, we see the need to perform a special observation campaign just to test the instrumental polarization of our setup.

\begin{table}
      \caption[]{OJ287 Observational Data 2012 November -- 2013 April}
  \centering
         \label{tab1}
     {\scriptsize
         \begin{tabular}{c c c c c c c}

    \hline \hline  
\medskip 
        Object & JD-2456200 & images  &exp &Total integration time  \\ [0.5ex] 
               &    (d)             &   & (sec)&(sec) \\ [0.5ex] 
\hline
HD10476 & 49.3960 &  30 &0.2  &6  \\
\hline
HD14433 & 49.4071 &  30 &0.5  &15  \\
\hline
OJ287 &  49.4946 &  30 &60   &1800   \\
OJ287 &  49.5233 &  30 &60   &1800   \\
OJ287 &  49.5521 &  30 &60   &1800   \\
OJ287 &  49.5878 &  30 &60   &1800   \\
OJ287 &  49.6176 &  30 &60   &1800   \\
OJ287 &  49.6465 &  20 &60   &1200   \\
\hline	
OJ287 &  50.4909 &  10 &60  &600    \\
OJ287 &  50.5003 &  10 &60  &600    \\
OJ287 &  50.5097 &  10 &60  &600    \\
\hline 
HD90508 & 106.4322 & 10 &3  &30  \\
HD90508 & 106.4354 & 10 &1  &10  \\
\hline
HD43384 & 106.4541 &  10 &1  &10  \\
\hline
OJ287 &  106.3641 &  10 &300   &3000   \\
OJ287 &  106.4956 &  10 &300   &3000   \\
\hline
HD144287 & 187.4698 & 5 &0.5  &2.5  \\
HD144287 & 188.4083 & 5 &0.5  &2.5  \\
\hline
HD154445 & 187.4837 &  5 &0.1 &0.5  \\
HD154445 & 188.4776 &  5 &0.2  &1  \\
\hline
OJ287 &  187.4481 &  3 &300   &900   \\
\hline
OJ287 &  188.3593 &  15 &300   &4500   \\
\hline
\end{tabular} }
\end{table}	

\begin{table*}
      \caption[]{Standard Stars Characteristics}
  \centering
         \label{tab2}
     {\scriptsize
         \begin{tabular}{c c c c c c c c}
    \hline\hline
\medskip
       Star & R. A. (2000) & Dec. (2000) & mV   & Polarization (max) & P.A.(cat)& P.A.(measured + error)&P.A.(measured)+P.A.(cat) \\ [0.5ex] 
            & hh mm ss     & dd '' ````  & mag  & ($\%$)             & ($deg$)& ($deg$)& ($deg$, corrected for 360 $deg$ rotation)   \\ [0.5ex] 
\hline
HD10476 & 01 42 29.8 &  +20 16 07& 5.2 & - & -& - & -  \\
\hline
HD14433  & 02 21 55.4 &  +57 14 34  & 6.39 & 3.9 at lmax=0.51$\mu$m & 112& 355.53 ($\pm$ 2.52) &  107.53($\pm$ 3.01) \\
\hline 
HD90508 & 10 28 03.9 &  +48 47 06 & 6.44 & - & -& - & -  \\
\hline
HD43384 & 06 16 58.7 &  +23 44 27 & 6.29 & 3.0 at lmax=0.53$\mu$m & 170& 333.18($\pm$ 2.18) & 143.18($\pm$ 2.68) \\
\hline
HD144287 & 16 04 03.7 &  +25 15 17 & 7.06 & - & -& - & -  \\
\hline
HD154445 & 17 05 32.3 &  $-$00 53 31  & 5.64 & 3.7 at lmax=0.57$\mu$m & 90& 65.72 ($\pm$ 0.76) & 155.72 ($\pm$ 1.26) \\
\hline
\end{tabular} }
\end{table*}	

\section{Results and Discussion}
Results for the polarization and position angle are shown in Table \ref{tab3}. Photometrical measurements are shown in Table \ref{tab4}. 

In Table \ref{tab5} we present the standard stars measured polarization and the catalogue values. The discrepancy accounts for the intrinsic instrumental polarization. The errors are pretty large, but still, given the number of observing nights and weighting  all the observations, we can get relatively good approximations. In addition, note that the observed optical polarized flux is contaminated by unpolarized flux and this strongly affects the polarization and P.A. determination using Stokes parameters, as explained in Villforth et al. (2010).

OJ287 polarization and P.A. are known to vary extremely rapidly, even in a matter of hours (Takalo et al. 1994; D'arcangelo et al. 2009). Polarimetrical and P.A. measurements from November 2012 (Bozhilov et al. 2013) indicated an agreement with previous data (Takalo, Sillanp\"{a}\"{a} \& Nilsson. 1994; Efimov et al. 2002; Villforth et al. 2010).

Efimov et al. (2002) argue that the plane of polarization angle rotates at a rate of about 5 $deg$ per day. Our preliminary 2-epoch results (Bozhilov et al. 2013) showed a change of 10.8 $deg$ per day. Considering the observation error limits of that preliminary analysis, we concluded that our data are not in contradiction with previous measurements (Efimov et al. 2002). 
But when we analyze the full data set, as presented on Table \ref{tab3} and Table \ref{tab4}, we can further improve on our previous work. 

We can represent the data in the column P.A. in Table \ref{tab3} using: 
\begin{equation}
  \label{eq1}
  P.A.=\omega*dT-k*360\ 
\end{equation}

 where
 $P.A.$ is the observed P.A.,
 $\omega$            is the change of P.A. in $deg$ per day,
 $dT$    is the difference in time between each two consecutive observational nights,
 $k$  is the number of complete turns of the plane of polarization.

That is done in order to determine the rotation of the plane of polarization and to account for the number of complete circles of the plane. This is the meaning of column P.A.(k) in Table \ref{tab3}. Some details are important regarding the determination of the parameter $k$. Here we assume that one full circle of the plane of polarization is somewhere between 5 $deg$ per day (value according to Efimov et al. 2002) and no more than 10.8 $deg$ per day (according to our previous work in Bozhilov et al. 2013). Thus, for the 81 days betwen the observations on the night of 2013 January 13 and the night of 2013 Arpil 04, we conclude that the plane of polarization could have made one additional circle. This, combined with the inherent 180 $deg$ ambiguity of the electric vector P.A. measurement, motivates our choice of k = 2.5 in Table \ref{tab3}. Note that we have made tests with different values of k (e.g. k= 2,3) with different rotation of the polarization plane, but the fit to observational data was not so good. 

\begin{table}
      \caption[]{Polarization and P.A. Measurements for OJ287} 
  \centering
         \label{tab3}
     {\scriptsize
         \begin{tabular}{c c c c c c c}
    \hline\hline
\medskip
      JD-2456200 & Polarization & Error & P.A. & Error & k & P.A.(k)\\ [0.5ex] 
        &  $\%$ & $\%$ &  $deg$ & $deg$ & & $deg$ \\ [0.5ex] 
\hline
49.4946 & 9.80 & 2.72 &  73.29  & 3.15 & 0&   73.29  \\
49.5233 & 9.84 & 2.69 &  74.52  & 3.15 & 0&   74.52  \\
49.5521 & 9.71 & 2.67 &  74.77  & 3.15 & 0&   74.77  \\
49.5878 & 9.89 & 2.76 &  74.62  & 3.15 & 0&   74.62  \\
49.6176 & 9.45 & 2.78 &  73.53  & 3.16 & 0&   73.53  \\
49.6465 & 9.48 & 2.76 &  72.53  & 3.16 & 0&   72.53  \\
\hline					      
50.4909 & 9.27 & 6.56 &  62.60  & 3.37 & 0&   62.60  \\
50.5003 & 9.54 & 5.31 &  64.45  & 3.29 & 0&   64.45  \\
50.5097 & 9.53 & 5.34 &  64.17  & 3.29 & 0&   64.17  \\
\hline %
106.364 & 4.93 & 5.10  & 117.33 & 3.20 & 1&   -242.67 \\
106.496 & 5.07 & 3.91  & 117.91 & 3.07 & 1&   -242.09  \\
\hline	        			      
187.448 & 25.59& 13.90 & 163.65 & 1.53 & 2.5& -736.35 \\
\hline					      
188.348 & 19.37 &10.52 & 172.35 & 1.53 & 2.5& -727.66 \\
\hline	   	                  
188.385 & 19.35 &9.91  & 172.37 & 1.52 & 2.5& -727.63 \\
\hline
\end{tabular} }
\end{table}	

\begin{table}
      \caption[]{Photometry of OJ287 }
  \centering
         \label{tab4}
     {\scriptsize
         \begin{tabular}{c c c }
    \hline\hline
\medskip
       JD-2456200 & Magnitude (R) & Error  \\ [0.5ex] 
        &   &     \\ [0.5ex] 
\hline
49.4655  & 14.614  & 0.006 \\
49.4659  & 14.606  & 0.006 \\
49.4664  & 14.614  & 0.006 \\
49.4676  & 14.604  & 0.006 \\
49.4681  & 14.610  & 0.005 \\
49.4685  & 14.603  & 0.005 \\
49.4715  & 14.602  & 0.005 \\
49.4725  & 14.603  & 0.005 \\
49.5694  & 14.599  & 0.005 \\
49.5700  & 14.602  & 0.004 \\
49.5705  & 14.598  & 0.004 \\
49.6536  & 14.611  & 0.005 \\
49.6541  & 14.619  & 0.006 \\
49.6545  & 14.617  & 0.005 \\
\hline           	   
50.4812  & 14.656  & 0.008 \\
50.4816  & 14.662  & 0.008 \\
50.4821  & 14.663  & 0.007 \\
\hline	
106.3364 & 15.413  & 0.005 \\
106.3399 & 15.412  & 0.005 \\
106.3427 & 15.420  & 0.005 \\
187.4144 & 14.667  & 0.003 \\
187.4206 & 14.666  & 0.003 \\
187.4223 & 14.666  & 0.003 \\			   
\hline	 
188.4181 & 14.707  & 0.004 \\
188.4197 & 14.708  & 0.003 \\
188.3279 & 14.711  & 0.003 \\
188.3295 & 14.713  & 0.003 \\
188.3311 & 14.719  & 0.004 \\
\hline 
\end{tabular} }
\end{table}	

\begin{table*}
      \caption[]{Standard Stars Catalogue and Observed Polarization} 
  \centering
         \label{tab5}
     {\scriptsize
         \begin{tabular}{c c c c c c}
    \hline\hline
\medskip
       Star & Polarization (max) &JD & Polarization & Error   \\ [0.5ex] 
           & ($\%$)  & &($\%$) & ($\%$) \\ [0.5ex] 
\hline
HD10476 &  - & 2456249.39580 &1.19 & 9.89  \\
\hline
HD14433  & 3.9 at lmax=0.51$\mu$m & 2456249.40664& 2.34 & 11.89\\
\hline 
HD90508  & - & 2456306.4312037& 3.37 & 6.52 \\
\hline
HD43384  & 3.0 at lmax=0.53$\mu$m &2456306.454097& 3.13 & 13.51\\
\hline
HD144287  & - & 2456387.4693056 & 19.78 & 17.99  \\
\hline
HD154445 & 3.7 at lmax=0.57$\mu$m & 2456387.483692 &20.18 & 30.77 \\
\hline
\end{tabular} }
\end{table*}	

 \begin{figure*} 
      \centering
      \includegraphics[width=120mm,height=62mm]{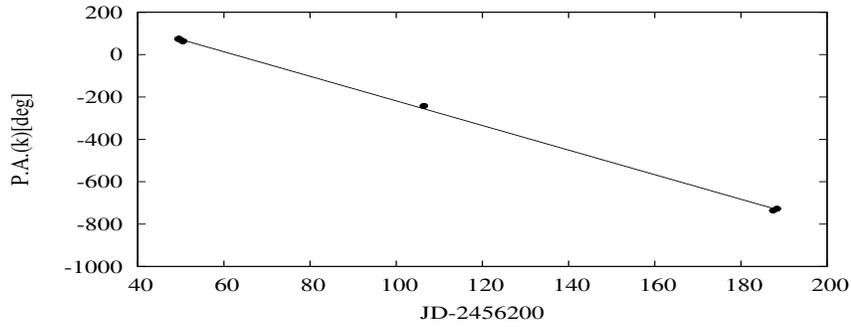}
      \caption{
P.A. (k) for OJ287 from Table \ref{tab3}. Observed P.A. change corresponds to rotation of 5.80($\pm$ 0.03) $deg$ per day.
              }
      \label{fig2}
   \end{figure*}

 \begin{figure*} 
      \centering
      \includegraphics[width=140mm,height=82mm]{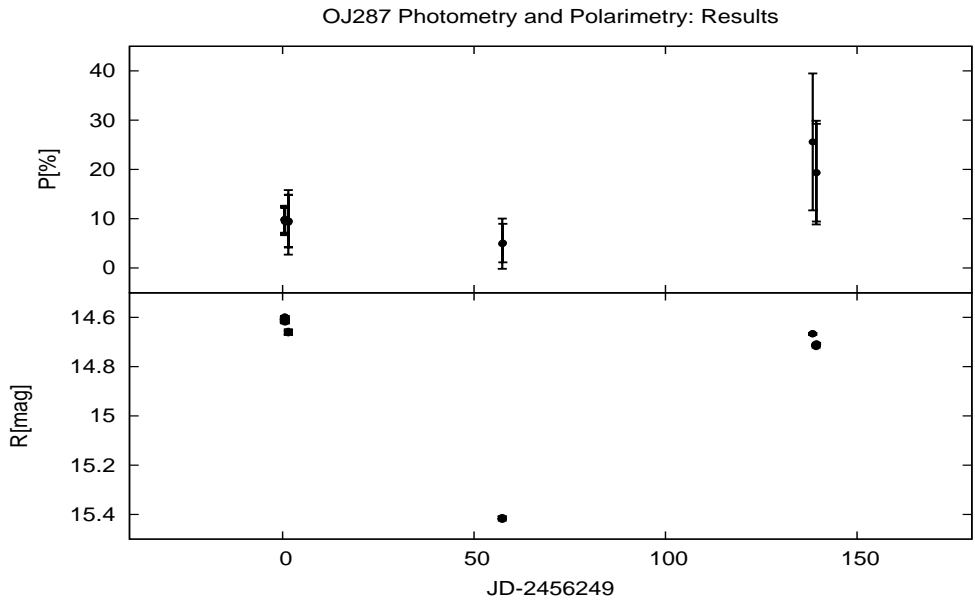}
      \caption{
Photopolarimetry for OJ287 based on the observational data from Table \ref{tab3} and Table \ref{tab4}.
              }
      \label{fig3}
   \end{figure*}

We derive P.A. change of 5.80 ($\pm$ 0.03) $deg$ per day. Results are plotted on Figure \ref{fig2}. This is closer to the observed change in Efimov et al. (2002). Note that this change in P.A. is likely due to underlying changes in the magnetic field of the structure (Homan et al. 2002). Efimov et al. (2002) and Valtonen \& Pihajoki (2013) attribute the rotation of the plane of polarization to the helical structure of magnetic field of the jet in OJ287.

The change of polarization versus photometric variability is shown on Figure \ref{fig3}. Efimov et al. (2002) did not found siginificant results to establish a correlation between change in brigthness and polarization variation. Jorstad et al. (2007) and Villforth et al. (2010) got to the same conclusion. But D'arcangelo et al. (2009) showed that there is a clear tendence for higher polarization when the optical flux is stronger e.g. when the object is getting brighter. This is an expected theoretical result in some models (Valtonen \& Sillanp\"{a}\"{a} 2011). In our results, we also observe such behaviour. On Figure \ref{fig3} you can see the photometrical and polarization change during our observational campaign. Our data are in concordance with some previous work by other authors (D'arcangelo et al. 2009) and support the expected photopolarimetrical dependence by Valtonen \& Sillanp\"{a}\"{a} (2011).

\section{Conclusions}

The Blazar-type AGN OJ287 is one of the most well-observed blazars. Yet, it still remains a riddle. Numerous models have been proposed to explain its observed properties (Valtonen \& Ciprini 2012). Since optical polarization measurements could be useful to distinguish among various theoretical models, our observations further complete the existing data on this peculiar object. Our original 5-epoch study of the optical photometry, polarization and P.A. of OJ287 shows good correlation with previous data  and gives strong background to further measurements with NAO Rozhen's focal reducer FoReRo2 on the 2-m RCC telescope. We observe variation in P.A. that corresponds to rotation of the plane of polarization of 5.80 $deg$ per day. Thus, our work builds on and impoves recent previous research on OJ287 (D'arcangelo et al. 2009; Bozhilov et al. 2013). We also observe dependence of the polarization behaviour with change of optical brightness, as expected by the Binary Black Hole model (Valtonen \& Sillanp\"{a}\"{a} 2011).

\section*{Acknowledgments}

The authors would like to thank Dr. Petko Nedialkov from the Department of Astronomy, Faculty of Physics, Sofia University, as well as Dr. Galin Borisov and Dr. Tanyu Bonev from the Institute of Astronomy for valuable advices and guidance. The authors gratefully acknowledge observing grant support from the Institute of Astronomy and Rozhen National Astronomical Observatory, Bulgarian Academy of Sciences.

\newpage


\label{lastpage}

\end{document}